\documentclass{article}[14pt]
\usepackage{epsfig}
\usepackage{bm}
\usepackage{colordvi}

\begin{document}
\title{					
\large{\bf 
Dynamical Likelihood Method 
for Reconstruction  
of Quantum Process \vspace*{1cm}}}
\author{Kunitaka Kondo \vspace*{4mm}\\ 
{\it Advanced Research Institute for Science and Engineering},\\
{\it Waseda University, Tokyo 169-8555, Japan}}
\maketitle

\begin{abstract} 
The dynamical likelihood method for analysis of 
high energy collider events is reformulated. 
The method is to reconstruct the elementary parton state  
from observed quantities. 
The basic assumption is that each of final state partons occupies a unit 
phase space. 
The parton kinematics is statistically reconstructed using  
(a) virtual masses of resonant partons and (b) parton kinematic 
quantities inferred  from observed quantities.  Generation of (b) is made 
with the transfer function which is the probability function for  parton 
kinematics from a given set of observables. 
Corresponding to the unit parton phase space, the transfer variable 
spaces are also quantized.
The likelihood of the reconstructed state  is 
defined by the Poisson probability for a single event with the expected 
number of event that is the cross section per unit phase space times 
a luminosity factor.
Applications of the method to selection of process, parton-observable  
identification, determinations of  parton 
kinematics and dynamical parameters are discussed.
\end{abstract}

\clearpage
\section{Introduction}
Dynamical Likelihood Method (DLM) was originally proposed in Ref. \cite{kk1} 
and developed  in Refs. \cite{kk2} and \cite{kk3} as a method to determine 
dynamical 
parameters, e.g. masses, decay widths or coupling constants 
from measured quantities ({\it observables}).
The basic idea is to use the differential cross section ( d.c.s. ) as 
a theoretical input for the event reconstruction.
A formulation of the method to use the d.c.s. as a posterior probability 
was given in Ref. \cite{dg}, and was used in Ref. \cite{d0}.

An alternative formulation of the posterior probability with the d.c.s.  
is presented in this paper. The motivation is as follows. 

In the traditional use of 
the d.c.s., it is defined per certain kinematic quantities and integrated 
by other kinematic variables. This is to match the detector arrangements 
devised for individual experiments. The d.c.s. there is the prior probability 
and can be essentially applied to a large number of events.
In the general purpose collider 
experiments, however, the detectors are not designed to measure particular 
quantitites but to make it possible to get over-all picture of each event. 
Hence the event reconstruction on event by event basis is feasible. 
To deal with a single event, we postulate that partons 
in the final state of hard scattering occupy a unit phase space. Thus 
the d.c.s. is defined per unit phase space of the final partons. 

Leptons, quarks and gauge bosons are called {\it partons} in this paper.
Parton process described by the Feynman diagram, i.e. a hard scatterring 
process, is called the {\it elementary} or {\it parton level} process.
A process from the initial beam state to the observables is called 
a {\it path}.  A path that a real event has taken is unknown. The event 
reconstruction is to infer paths which the event could have taken. 

The {\it probability density function} ( p.d.f. ) of the first half a path, 
i.e. from the beam to the parton state,   is defined by the d.c.s. per unit 
phase space ( quantum state ) of the final partons in the elementary process. 
This is consistent with the original definition of the d.c.s.

The p.d.f. for the second half a path, i.e. from the parton state to 
observables,  is the transfer function ( T.F. ) that  relates 
the observables with the corresponding variables at the parton level.  
Corresponding to the quantum condition on the final parton state, 
the transfer variable spaces are also quantized. 

The third quantum condition is about the number of events. 
The d.c.s. multiplied with a luminosity factor can be interpreted as 
an expectation value of 
the number of events, and the likelihood ( posterior probability ) of the 
event is defined by the Poisson probability for 1 event 
with the given expectation value. 

 
This paper is organized as follows: 

In Sec. $\bm{2}$, the d.c.s. for the final parton state and its use 
for the likelihood definition are discussed.  The parton level likelihood 
works as the likelihood of the reconstructed parton process as we see in 
the following sections. 

The subject of Sec. $\bm{3}$ is the T.F. 
A way to obtain the T.F. with fully simulated Monte Carlo events is proposed. 
The detection efficiency associated with measurements and 
event selection criteria is automatically included in the T.F. 
According to the quantum parton state, the Jacobian scaled transfer 
variables are introduced.    As a result, 
the likelihood is essentially Jacobian free. The T.F. in its posterior 
form is used for parton level reconstruction.

An efficient way of inferring paths is discussed in Sec. $\bm{4}$. 
Momenta of observed final partons are inferred with the posterior T.F. 
using observed quantities as inputs. 
Virtual masses squared of resonant primary partons are 
inferred with their propagator factors. 

In Sec. $\bm{5}$, the likelihood for reconstructed multiple parton states is 
discussed.   To regenerate the unknown true path of a given event, one makes 
multiple inference of paths in the event. 
For the multiple inferences, the maximum, 
the expectation and the multiplicative values of the likelihood  
are defined. They are used  statistically to select   
the process, the parton-observable identification and the kinematic 
solution of the secondary partons.
The determination of the dynamical parameters is made by the joint 
likelihood of all events in the sample.  
 
Summary of the formulation is given in Sec. $\bm{6}$.

\section{Differential Cross  Section and Parton Level Likelihood}
In DLM, the d.c.s. for an inferred parton process is used 
as a theoretical input to evaluate the likelihood of the reconstruction. 
\subsection{Differential  cross section for a final parton state}
We assume that a process is described at the parton level by 
\begin{equation}
\label{eq:process}
  a/A+b/B  \rightarrow \cdots \rightarrow c_{1}+\cdots + c_{n} \equiv C,
\nonumber
\end{equation}
where $a$ and $b$ are the initial partons, each representing a quark or 
an anti-quark or a gluon,  in beam particles $A$ and $B$ respectively, and
$c(c_{1},c_{2},...c_{n})$ are the final state partons. 
States of partons are  after the initial- and before the final-state 
radiations. 
Throughout this paper, particle symbol $p$ also represents its 4-momentum, 
and $\bm{p}$ its 3-momentum.
The final partons are assumed to be on mass-shells, i.e. 3-momenta are enough 
to define their states. 
Process (\ref{eq:process}) as a whole, i.e. a set of momenta of all partons, 
is called {\it parton state} in this paper.

Beam particles $A$ and $B$ are assumed to make a head-on collision 
along the $z$-axis. Then the hadronic cross-section for process 
(\ref{eq:process}) is given by
\begin{equation}
\label{eq:dsigma}
d\sigma
= dz_{a}  dz_{b}  dP_{T}  
f_{a/A}(z_{a},\bm{\alpha}) f_{b/B}(z_{b},\bm{\alpha}) 
f_{T}(P_{T},\bm{\alpha}) d\hat{\sigma}
(a+b\rightarrow C;\bm{\alpha}),
\end{equation}
where  $d\hat{\sigma}$ is the parton level cross section,
\begin{equation}
\label{eq:dsighat}
d\hat{\sigma}(a+b \rightarrow C;\bm{\alpha}) =
\frac{(2\pi )^4 \delta^{4}(a+b-C)}{4\sqrt{(a\cdot b)^2 -m_a^2m_b^2}} 
| {\cal{M}}(a+b\rightarrow C;\bm{\alpha}) 
|^2 d\Phi^{(f)}_n.
\end{equation}

In Eq. (\ref{eq:dsigma}), symbol  $\bm{\alpha}$ stands for a set of dynamical 
constants, e.g.  masses, decay widths or coupling constants. Hereafter, 
we use symbol $\bm{\alpha}$ to represent only unknown parameters to 
be measured.   Variables $z_{a} = a_{z}/|\bm{A}|$ and 
$z_{b} = b_{z}/|\bm{B}|$ 
are momentum fractions of $a$ and $b$ 
in hadrons $A$ and $B$ respectively, 
and $P_{T}$ is the total momentum of 
the initial/final system of process (\ref{eq:process}) in the plane 
perpendicular to the beam axis.  
The probability density functions ( {\it p.d.f.}) for $z_{a}$, $z_{b}$ and 
$P_{T}$ are  denoted by $f_{a/A}$, $f_{b/B}$, and $f_{T}$, respectively.
Functions $f_{a/A}, f_{b/B}$ and $f_T$ are 
effective parton distribution functions for process (\ref{eq:process}) 
with the radiation parts removed. 
In practice,  these p.d.f.'s are to be obtained 
by running Monte Carlo event generators. 

In Eq. (\ref{eq:dsighat}), ${\cal{M}}$ is the matrix element for process 
(\ref{eq:process}), and $d\Phi^{(f)}_n$ is the differential factor,  
\begin{equation}
\label{eq:dphasen}
d\Phi^{(f)}_{n} \equiv \prod _{i=1} ^n \frac{d^3\bm{c}_i}{(2\pi)^3 2E_i},  
\end{equation}
of the Lorentz invariant $n$-body phase space element,
\begin{equation}
d\Phi_{n} = \delta^{4}(a+b-C) d\Phi^{(f)}_{n}.
\end{equation}
We call $d\Phi^{(f)}_{n}$ the {\it phase space element} ( P.S.E. ) 
in this paper. 

The d.c.s. for a final state defined by 
$\bm{c}(\bm{c}_{1},\cdots,\bm{c}_{n})$ is obtained by integrating 
Eq. (\ref{eq:dsigma}) with the initial state variables $z_{a}$, $z_{b}$ 
and $p_{T}$ as   
\begin{equation}
\label{eq:dsprdphi}
d\sigma 
= I(a,b) \left|{\cal{M}}(a+b\rightarrow C;\bm{\alpha})\right|^2  
d\Phi_{n}^{(f)} ,
\end{equation}
namely, 
\begin{equation}
\frac{d\sigma}{d\Phi_{n}^{(f)}}
= I(a,b) \left|{\cal{M}}(a+b\rightarrow C;\bm{\alpha})\right|^2,
\label{eq:dsdfi}
\end{equation}
where
\begin{equation}
I(a,b) = \frac{(2\pi)^4}{4|\bm{A}||\bm{B}|\sqrt{(a\cdot b)^2 -m_a^2m_b^2}} 
f_{a/A}(z_{a},\bm{\alpha}) f_{b/B}(z_{b},\bm{\alpha}) 
f_{T}(p_{T},\bm{\alpha}).
\end{equation} 
The formulation in this paper is to define the likelihood to be proportional 
to $d\sigma/d\Phi^{(f)}_{n}$.

\subsection{Resonances ( internal lines )}
Assume a process where {\it resonance}  $r$, which corresponds to an internal 
line in the Feynman diagram, is produced as
\begin{equation}
\label{eq:prodr}
a+b \rightarrow r+c_{j+1}+c_{j+2}+\cdots +c_n,
\end{equation}
and subsequently decays into channel $\rho$ as
\begin{equation}
\label{eq:decr}
r \rightarrow \rho: c_{1}+c_{2}+\cdots +c_{j}. 
\end{equation}


\underline{\it Propagator factor}
$\:\:$
If the matrix elements for processes (\ref{eq:prodr}) and 
(\ref{eq:decr}) are given as ${\cal{M}}_{prod}$ and ${\cal{M}}_{dec}$ 
respectively,   
the matrix element squared for process (\ref{eq:process}) is factorized as
\begin{equation}
\label{eq:matprp}
|{\cal{M}}(a+b \rightarrow r+c_{j+1}+\cdots+c_{n} \rightarrow C;
\bm{\alpha})|^{2} =|{\cal{M}}_{prod}|^2 \Pi (s_{r})|{\cal{M}}_{dec}|^2 ,
\end{equation}
where $s_{r}$ is the virtual mass squared of resonance $r$, given by
\begin{equation}
s_{r}\equiv r^{2}=( \sum_{i=1}^{j} c_{i} )^{2}.  
\label{eq:qsq}
\end{equation}
The lowest order approximation for propagator factor $\Pi(s_{r})$ is given by 
the Breit-Wigner form, 
\begin{equation}
\label{eq:breit-wigner}
\Pi(s_{r})=
\frac{1} {(s_{r}-M^{2}_{r})^2 + M_{r}^{2}\Gamma^{2}_{r}},
\end{equation}
where $\Gamma_{r}$ is the total decay width of $r$. 
Improved forms taking the higher order effects into account are 
discussed in Ref \cite{pythia}.  

In the event reconstruction, if all $c_{i}$'s are inferred from observables, 
$s_{r}$ is given by Eq. (\ref{eq:qsq}), while if $s_{r}$ is generated 
according to the propagator factor, 
Eq. (\ref{eq:breit-wigner}), Eq. (\ref{eq:qsq}) gives 
a constraint to $(c_{1},\cdots,c_{j}$. 

\subsection{Parton level likelihood}

Reconstruction of a parton level process by using $s_{r}$'s according to 
the propagator factors and $c_{j}$'s with the transfer function will be 
discussed in the following Sections. In this subsection, we define the 
parton level likelihood assuming that a path, i.e. a set of parton kinematics 
$P(P_{1},\cdots,P_{N})$, where $N$ is the total number of partons in the 
process, is given. 

If parton kinematics  is given, the differential 
cross section $\varsigma \equiv d\sigma/d\Phi_{n}^{(f)}$ can be caluculated. 
The expected number of events with cross section $\varsigma$ is
\begin{equation}
\label{eq:mu}
\mu=l_{1}\varsigma\equiv l_{1}\frac{d\sigma}{d\Phi_{n}^{(f)}},
\end{equation} 
where $l_{1}$ stands for a luminosity factor independent of kinematics of 
the path. We assume  that 
an event takes place according to the Poisson probability of a single event
$P(1;\mu)$, i.e. 
\begin{equation}
\label{eq:dpmu}
dP(1;\mu)=P(1;\mu)d\mu = P(\varsigma)d\varsigma,
\end{equation} 
and define the likelihood of the path by 
\begin{equation}
\label{eq:likemu}
L_{1}^{(0)}\equiv P(1;\mu) = \mu\;exp(-\mu),
\end{equation}
where suffix $1$ and prefix $(0)$ stand for a single path and the parton 
level, respectively. 
The p.d.f. for  
$\varsigma \equiv d\sigma/d\Phi_{n}^{(0)}$ is given by 
\begin{equation}
P(\varsigma)=l_{1}L_{1}^{0}=l_{1}\mu\:exp(-\mu)
=l_{1}^{2}\;\varsigma\; exp(-l_{1} \varsigma),
\end{equation}
from Eqs. (\ref{eq:mu}),(\ref{eq:dpmu}) and (\ref{eq:likemu}).

Events in a data set are mutually independent, hence the number of event 
distribution in a data set with the total number of events 
$N_{tot}$  is given as  
\begin{equation}
\frac{dN}{d\mu}=N_{tot} L_{1}^{(0)}.
\end{equation}

\underline{\it Peak value and the normalization of $L_{1}^{(0)}$}
$\;\;$
Likelihood $L_{1}^{(0)}$ takes its maximum value $1/e$ for $\mu=1$, since
\begin{equation}
\frac{dL_{1}^{(0)}}{d\mu}=(1-\mu)\;exp(-\mu),
\end{equation}
and it is normalized as 
\begin{equation}
\int_{0}^{\infty}L_{1}^{(0)}d\mu
=\int_{0}^{\infty}\mu \;exp(-\mu)d\mu=1.
\end{equation}

\underline{\it Luminosity factor $l_{1}$}$\:\:$
The expectation value of $\varsigma=d\sigma/d\Phi_{n}^{(f)}$ is
\begin{equation}
\left< \frac{d\sigma}{d\Phi_{n}^{(f)}} \right> 
=\int_{0}^{\infty} \varsigma \: P(\varsigma) d\varsigma 
=\frac{1}{l_{1}}\int_{0}^{\infty}\mu^{2}\: exp(-\mu) d\mu
= \frac{2}{l_{1}}.
\end{equation}
or
\begin{equation}
l_{1}=2 \left< \frac{d\sigma}{d\Phi_{n}^{(f)}} \right>^{-1}.
\end{equation}
If the integrated luminosity of the data set is $L_{int}$, then 
\begin{equation}
N_{tot}=L_{int}\left< \frac{d\sigma}{d\Phi_{n}^{(f)}} \right>
=\frac{2L_{int}}{l_{1}}.
\end{equation}
assuming the detection efficiency is 1. Then,  
\begin{equation}
l_{1}=\frac{2L_{int}}{N_{tot}}=\frac{2}{\sigma_{T}},
\end{equation}
where $\sigma_{T}$ is the total cross section for the process. In event 
reconstruction it may be interpreted as a function of the mass of particle 
in search, but since $L_{int}$ and $N_{tot}$ are measured/measurable 
quantities, we interpret them as observables  which are intrinsic to 
the given data sample just as the observed kinematic variables are. 
Formulation including the detection efficiency will be discussed in the 
later section.

\section{Transfer function}

The p.d.f. for the second half a 
path, i.e. a path from the parton state to the  
observables, is the transfer function ( T.F. ). 

\subsection{Observables in collider experiments}
Typical collider detectors have calorimeters and the tracking system. 
Calorimeters  and trackers with tgive energies and momenta of  
particles respectively. 
A jet is generally identified with a quark (anti-quark) or a gluon. 
An electromagnetic shower associated with or without a track 
is assigned to an electron or a photon. A track passing through  
calorimeters with a minimum ionizing signal is identified with a muon. 
We call these particles {\it observable partons} and their   
measured quantities {\it observables}.

\underline{\it Electrons, muons and photons}$\:\:$
These particles are relatively well identified and their momenta are 
measured within the detector resolutions. 

\underline{\it Jets}$\:\:$
Jets  are assigned to quarks or gluons. 
Measured quantities of jets have uncertainties due to 
statistical nature of parton shower, hadronization,
resolution of detectors and jet reconstruction algorithm.

As for the assignment of jets to partons and the relation 
between their momenta, we make following comments: 

(1) There are color flows between these partons, and the fragmentation is  
not independent among them. But the effect can be integrated 
in the transfer function to be discussed later.

(2) It is possible that a quark/gluon is observed as two or more jets. 
If two or more nearby jets are merged into one and identified
with a single parton successfully, the effect can be remedied. 
Otherwise, the effect results as an inefficiency of the reconstruction 
and/or a shift in the values of dynamical parameters to be determined.

(3) In general 4-momentum $(E_{J},\bm{p}_{J})$ is measured on a jet. 
For quarks in the final partons, however, we assume their pole masses.  
Hence 3 quantities of a jet are enough to infer the quark 3-momentum. 
The selection of these quantities is not unique  but is to be made 
according to the process, the purpose of analysis and detector properties. 

\underline{\it Missing partons}$\:\:$
For partons which do not interact with 
detectors, e.g. neutrinos,  the missing transverse energy(MET)   
$\bm{\not \!\!E}_T$ defined by
\begin{equation}
\label{eq:met}
 \bm{\not \!\!E}_T=-\bm{E}_{T}^{(obs)}
                  =-(\bm{E}_{T}^{(cal)}+\sum \bm{\mu}_{T})
\end{equation}
is measured, where $\bm{E}^{(obs)}_{T}$ is the  
measured total transverse energy flow, $\bm{E}_{T}^{(cal)}$ is the sum of the 
transverse energy flow measured by calorimeters, and $\sum \bm{\mu}_{T}$ is   
the sum of  transverse momenta of muons measured 
by the tracking detector. All vectors in Eq. (\ref{eq:met}) are 
in the plane perpendicular to the beam-axis.

\subsection{Transfer functions for observable quantities}

For a real event in experiment, the final observables are known, 
while the parton state in process (\ref{eq:process}) is unknown. 
The event reconstruction in DLM is to infer such unknown parton  
state that  leads to an observed variable set 
$\bm{y}(y_{1},\cdots,y_{N_{V}})$. 
The parton variable set corresponding to $\bm{y}$  is denoted by  
$\bm{x}(x_{1},\cdots,x_{N_{V}})$.
We call $\bm{x}$ and $\bm{y}$  {\it transfer variables}.

\underline{\it Prior transfer function}$\:\:$
The prior {\it transfer function} (T.F.) is a p.d.f. for $\bm{y}$ 
when $\bm{x}$ is given and denoted by 
$w(\bm{y}|\bm{x}|| i_{p}, \bm{\alpha})$, 
where $i_{p}$ is an integer to specify the process. 
The probability for $(\bm{x}, \bm{y})$ to be in $(d\bm{x}, d\bm{y})$ is 
\begin{equation}
\label{eq:wdxdy}
dP(\bm{x}, \bm{y}||i_{p})  
= w(\bm{y}|\bm{x}||i_{p},\bm{\alpha})\, d\bm{x}d\bm{y}.
\end{equation} 
where
\begin{equation}
d\bm{x} \equiv \prod_{m=1}^{N_{V}} dx_{m},\quad
d\bm{y} \equiv \prod_{m=1}^{N_{V}} dy_{m}. 
\end{equation}

If $w(\bm{y}|\bm{x}||\bm{\alpha})>0$, a certain value of $\bm{y}$ should
exist.   
Hence we require the normalization condition,
\begin{equation}
\label{eq:wnorm}
\int_{w>0} w(\bm{y}|\bm{x}||\bm{\alpha}) d\bm{y} = 1,  
\end{equation}
for any $\bm{x}$ with $w>0$. 

A typical example of a component of $\bm{y}$ is the energy of a jet. 
The  T.F. for a jet depends on models of parton-shower and fragmentation, 
the detector response and the jet reconstruction algorithm.  
Thus it is appropriate to derive the T.F. by using Monte Carlo event 
generators with full simulation, where the momentum of each parton 
and measured quantities associated with it are provided. 
Events are to be selected with the same criteria as applied to 
real data.

\underline{\it  $w(y|x||\alpha)$ from Monte Carlo events}$\:\:$
The T.F. is a function of multi-dimensional variables 
$\bm{x}$ and $\bm{y}$. 
We assume that the T.F. is factorisable as 
\begin{eqnarray}
 w(\bm{y}|\bm{x}||i_{p},\bm{\alpha})
=\prod_{m=1}^{N_{V}} w(y_{m}|\bm{x}||i_{p},\bm{\alpha}).
\label{eq:wfact}
\end{eqnarray}

To illustrate how to get T.F., we take a simple case where a T.F. 
for $y_{m}$ depends on the corresponding variable $x_{m}$ only, and 
$\bm{\alpha}$ has a single component $\alpha$. 
We denote the  T.F. by $w(y|x||\alpha)$, 
abbreviating process number $i_{p}$ and variable number $m$. 

Let $n_{xy}$ denote the density of generated number of events at $(x,y)$, 
and $n_{x}$ that at $x$. 
T.F. $w(y|x||\alpha)$ is defined such that the number of events 
in $(dx,dy)$ is given by
\begin{equation}
dN(x,y) = n_{xy}dx dy = n_{x}dx \times w(y|x||\alpha)dy.
\label{eq:dn}
\end{equation}
With the integrated luminosity $L_{int}$, the number densities are given by
\begin{equation}
\label{eq:nxnxy}
n_{xy} 
=  L_{int} \frac{d\sigma}{dx} \, w(y|x||\alpha),
\:\:\:
n_{x} =\int n_{xy}dy  = L_{int} \frac{d\sigma}{dx}.
\end{equation}
The $y$ dependence of the detection efficiency is included in 
$w(y|x||\alpha)$.   

From Eq. (\ref{eq:dn}), the T.F. is given by
\begin{equation}
\label{eq:priw}
w(y|x||\alpha) = \frac {n_{xy}}{n_{x}}
=\sum_{i=1}^{n_{xy}} \frac{1}{n_{x}}.
\end{equation}
Thus $w(y|x||\alpha)$ is obtained by filling the $(x,y)$ 
histogram with weight $1/n_{x}$ for each event. 
Weighting by $1/n_{x}$ is to avoid the double counting of the cross 
section factor which exists in the parton level likelihood.
Integrating Eq. (\ref{eq:priw}) by $y$ and using Eq. (\ref{eq:nxnxy}), one 
obtains the normalization condition, Eq. (\ref{eq:wnorm}). 
Note that the correction for the detection inefficiency associated 
with measurements and event selection conditions is automatically made 
by deriving the T.F. with the Monte Carlo events.

\subsection{Quantization of the transfer variable space}
We consider how the quantum condition $\Delta\Phi_{n}^{(f)}=1$ characterizes 
the transfer variable spaces. 
This condition applies to all final partons, while the transfer 
variables make sense only for observable partons.
Hence we first discuss the case of observable final partons.

\underline{\it Jacobian scaled variables for observable final parton}
$\:\:$
We denote transfer variables of the $l$-th observable final parton by 
$\bm{x}_{l}$  and the corresponding observables by $\bm{y}_{l}$ 
( $l=1,\cdots,N^{*}_{obs}$ ). Variable $\bm{x}_{l}$ is a 3-component 
function of $\bm{p}_{l}$, and generally $N^{*}_{obs} \le n$, where 
$n$ is the number of final state partons. 
For the $l$-th observable final parton, we introduce 
variables $(\bm{X}_{l}, \bm{X}_{l})$ by  
\begin{eqnarray}
\label{eq:dxstar}
d\bm{X}_{l}&\equiv& \prod_{k=1}^{3} dX_{lk}
=\frac{d\Phi_{1}^{(l)}}{d\bm{x}_{l}}d\bm{x}_{l}
= J_{xl}d\bm{x}_{l},\\
\label{eq:dystar}
d\bm{Y}_{l}&\equiv& \prod_{k=1}^{3} dY_{lk}
=\frac{d\Phi_{1}^{(l)}}{d\bm{y}_{l}}d\bm{y}_{l}
= J_{yl}d\bm{y}_{l},
\end{eqnarray}
where $J_{xl}$ and $J_{yl}$ are the phase space Jacobian factors, 
\begin{eqnarray}
J_{xl}
&\equiv&\frac{1}{(2\pi)^{3}2E_{l}}\left|\frac{\partial(p_{lx}, p_{ly}, p_{lz})}
{\partial(x_{l1}, y_{l2}, x_{l3})} \right|, \\
J_{yl} &\equiv& J_{xl} |_{x=y}. 
\end{eqnarray}
Obviously, the unit phase volume $\Delta \Phi_{1}^{(l)}=1$ corresponds to
the unit variable spaces
\begin{equation}
\label{eq:qpath1}
\Delta \bm{X}_{l}=1, \:\: \Delta \bm{Y}_{l}=1.
\end{equation}

A time-ordered path may be described as follows. 
A single path specifies a unit phase volume (cell) of final parton $l$,   
which one-to-one corresponds to a unit volume (cell) of 
$\bm{X}_{l}$, and picks up that of $\bm{Y}_{l}$  statistically 
according to  the T.F. 
In other words, by condition $\Delta \Phi_{1}^{(l)} = 1$, variable 
spaces    $\bm{x}_{l}$ and $\bm{y}_{l}$ are quantized.  
The elements of these spaces  become from real (continuous) to countable 
(discrete) 
almost-infinite numbers. The {\it width} of the quantized single path is 
$\Delta \bm{X}_{l}=\Delta \bm{Y}_{l}=\Delta \Phi_{1}^{(i)}=1$.

\underline{\it Transfer functions for Jacobian scaled variables}$\:\:$
We denote T.F. for Jacobian scaled variable ($\bm{X},\bm{Y}$) 
by $W(\bm{Y}|\bm{X}||\bm{\alpha})$. 
To compare the two T.F.'s, $w$ and $W$, we again treat a case of a 
single variable set $(x,y)$ and $(X,Y)$. 
The number of generated event in $(dX,dY)$ is expressed in terms of 
$W(Y|X||\alpha)$ as 
\begin{equation}
dN(X,Y)=L_{int}\frac{d\sigma}{dX}\,  
W(Y|X||\alpha)\,dX dY.
\label{eq:dndxdyjac}
\end{equation}
But $dN$ should be proportional to the outlet path width, i.e. 
\begin{equation}
\label{eq:dnxycomp}
dN(X,Y) = J_{y}dN(x, y).
\end{equation}
Comparing Eqs. (\ref{eq:dn}),(\ref{eq:nxnxy}), (\ref{eq:dndxdyjac}) 
and (\ref{eq:dnxycomp}), 
one gets a scale invariance of T.F.,
\begin{equation}
\label{eq:wscaleinv}
W(Y|X||\alpha)
=w(y|x||\alpha).
\end{equation}


\underline{\it Posterior T.F.}$\:\:$
Posterior T.F. $w(x|y||\alpha)$ for a single component set $(x,y)$ is given by
\begin{equation}
\label{eq:posw}
w(x|y||\alpha)=
\frac{w(y|x||\alpha)}
{\int_{w>0} w(y|x||\alpha)dx}
\end{equation} 
and is to be used to infer parton variable $x$ from observable $y$. The 
posterior T.F. for Jacobian scaled variables $W(X|Y||\alpha)$ is obtained 
by
\begin{equation}
W(X|Y||\alpha)=\frac{W(Y|X||\alpha)}{\int_{W>0} W(Y|X||\alpha)\;dX}.
\end{equation}  

For a given $Y$, the value of $X$ is to be inferred by the probability,
\begin{equation}
dP(X;Y)=W(X|Y)\;dX.
\end{equation}
Using the scale invariance, Eq. (\ref{eq:wscaleinv}), one gets
\begin{equation}
dP(X;Y)=\frac{J_{x}}{\tilde{J}_{x}}\;w(x|y||\alpha)dx,
\end{equation} 
where $\bar{J}_{x}$ is the mean value of $J_{x}$ defined by
\begin{equation}
\bar{J}_{x}=
\frac{\int_{w>0} J_{x}\;w(y|x||\alpha)dx}{ \int_{w>0} w(y|x||\alpha)dx}.
\end{equation}
The domint part of $w(x|y||\alpha)$ is symmetric with respect to $x-\bar{x}$, 
where
\begin{equation}
\bar{x}=\int x\;w(x|y||\alpha)dx.
\end{equation} 
Hence the effect of  $J_{x}-\bar{J}_{x}$ is cancelled out in the first order, 
and $J_{x} \approx \bar{J}_{x}$. In this approximation,
\begin{equation}
dP(X;Y) \approx w(x|y||\alpha) dx. 
\end{equation}
Thus the variable quantization is required only conceptually, and in practice 
one can use the posterior T.F.  $w(x|y||\alpha)$ instead of $W(X|Y||\alpha)$.

\underline{\it Missing final partons}$\:\:$
The only observable about missing partons are the missing transverse energy. 
MET. 
The sum of transverse momenta of missing particles, $\bm{T}(T_{x},T_{y})$,  
is inferred with T.F. for MET, $\:w(\bm{T}|\bm{\not \!\!E}_{T}||\bm{\alpha})$.
The parton level cross section can be written as
\begin{equation}
\label{eq:dsdfimet}
\frac{d\sigma}{d\Phi_{n}^{(f)}}=
\frac{d\sigma}{d\Phi_{n}^{(f)}} 
\delta (T_{x}-\sum_{m=1}^{M}c^{*}_{mx}) dT_{x}\:
\delta (T_{y}-\sum_{m=1}^{M}c^{*}_{my}) dT_{y},
\end{equation}
where $c^{*}_{ix,iy}(i=1,\cdots,M)$  are the $(x,y)$ components of missing 
partons. $\delta$-functions in Eq.(\ref{eq:dsdfimet}) give constraints, 
\begin{equation}
\sum_{m=1}^{M}c^{*}_{mx} = T_{x},\quad\quad\sum_{m=1}^{M}c^{*}_{my} = T_{y}.
\label{eq:txty}
\end{equation}
Since the quantization requirement for each missing final parton, 
$\Delta \Phi_{1}^{(f)} = 1$, is for 3-dimensional variables, the requirement 
is compatible with the 2-dimensional constraint, Eq. (\ref{eq:txty}). 
To summarize,  the MET constraint Eq. (\ref{eq:txty}) is free from the 
quantization condition, and the phase space of each  reconstructed parton, 
whether observable or missing, is to be taken as 1.  
The value of $d\sigma/d\Phi_{n}^{(f)}$ is evaluated with the transverse 
momentum components determined with constraint $(\ref{eq:dsdfimet})$.



\section{Path Reconstruction}

\subsection{Primary and secondary partons in event reconstruction}

DLM is a procedure to reconstruct the parton state,  
i.e. a set of momenta of all partons, $P(P_{1}, \cdots, P_{N})$,  
including  resonances and final partons. 
The parton kinematics is defined in general by giving momenta of $n$ 
out of $N$ partons. 
We call such $n$ partons the {\it primary} partons, 
denoting them by $p(p_{1}, \cdots, p_{n})$. 
Momenta of residual partons are determined by the energy-momentum 
conservation at vertices of the Feynman diagram. We call these partons 
{\it secondary} partons. 
These names are only to specify roles of partons in the event reconstruction.
The selection of the primary partons is optional, depending on the process 
and  the reconstruction algorithm.

\subsection{Specifications of process, topology and solution}

Given an event with observable set $\bm{y}$, there are 3 integers to 
specify a path. 

\underline{\it Process}$\:\:$
First, one has to assign 
physics process $i_{p} (i_{p}=1,\cdots,N_{p})$ which $\bm{y}$ came from. 

\underline{\it Topology}$\:\:$
Some of observed partons in an event cannot be uniquely identified 
with final partons in the elementary process.
Examples are the same sign electrons or muons, multiple photons or jets. 
In the event reconstruction, one has to assign 
some components of observable $\bm{y}$ to a set of parton species to define 
variable $\bm{x}$.
We call each set of the parton assignment to $\bm{y}$ a  {\it topology} 
in this paper, 
and denote the topology number by $i_{t}: i_{t}=1,\cdots,N_{t}$, where 
$N_{t}$ depends on process $i_{p}$.
Variable $\bm{x}$ and hence the value  of the T.F. 
depend on the assumed topology $i_{t}$. 

\underline{\it Solution}$\:\:$
If a process includes resonance(s), whether daughters are missing or 
observed, one can infer $\bm{s}_{r}$ and solve  Eq. (\ref{eq:qsq}) 
for momentum component(s) of daughter parton(s). 
The solutions are sorted by the solution number 
$i_{s}: i_{s}=1,\cdots,N_{s}$, where $N_{s}$ depends on $i_{p}$ and $i_{t}$.

\subsection{Outline of path reconstruction}
The  procedure of a single path  reconstruction is summarized below.\\ 
(1) One specifies process $i_{p}$  and infers $\bm{\alpha}$  uniformly,\\ 
(2) One specifies  topology $i_{t}$, and infers parton kinematics  
as follows:

(a)  One specifies an appropriate set of $n$ primary partons.
If all primary partons are observable, one infers their momenta 
$\bm{p}$ according to T.F. 
Totally missing partons are classified to primary partons, and a set of their 
momenta is to be inferred uniformly in their phase space. 

(b) If a resonance is assigned to a primary parton, one infers its 
invariant mass squared $\bm{s}_{r}$ with the propagator factor 
$\Pi(s_{r})$, and determines a secondary parton 
momentum by Eq. (\ref{eq:qsq}).  

Such inferences of variables $\bm{p}$ and $s_{r}$ are more efficient than 
scanning them uniformly. We call such inferences importance sampling 
( I.S. ). 

\subsection{ Inference of parton momentum from jet}
Quarks and gluons in the final parton state are observed as jets 
$(j_{1},\cdots,j_{N_{jet}})$. 
The parton momentum  can 
be inferred from observables of corresponding jet  by using T.F. 
$w(\bm{y}_{i}|\bm{x}_{i})$. 
In the following, we abbreviate parton/jet 
suffix $i$. Variable $\bm{x}$ can be  
$(E,\theta,\phi)$, $(E_{T},\eta,\phi)$ of the parton, or any other set as 
long as it is observable and determines $\bm{c}$ uniquely. 
 
An efficient way of inferring $x$, a component of $\bm{x}$,  is to make a 
variable transformation,
\begin{equation}
\label{eq:uniobs}
   u(x) = \frac{1}{x_{max}-x_{min}} 
\int_{x_{min}}^{x} w(y|\xi)d\xi,     
\end{equation}
where $u$ is a normalized uniform random number (n.u.r.n.: $0<u<1$), 
and range $(x_{min}, x_{max})$ is defined by $w>0$. 
Generating $u$, one can determine $x$. 

\subsection{ Inference of missing transverse energy}
We denote the transverse energy flow of the $i$-th missing parton by
$\bm{t}_{i}(t_{i} cos \phi_{i}, t_{i} sin \phi_{i})$. 
The total transverse energy of $m$ missing partons is
\begin{equation}
\label{eq:tmis}
\bm{T} = \sum_{i=1}^m \bm{t}_i.
\end{equation}
A simple example is a case where only one neutrino is involved in 
the process, where $\bm{T}=\bm{\nu}_T$. $\bm{T}$ is a parton variable 
to be inferred with the transfer function. 

The choice of the transfer variable set to infer $\bm{T}$ depends on 
whether process (\ref{eq:process}) includes partons going to jets  
or not.   
Let $\bm{x}_{T}$ and $\bm{y}_{T}$ denote such a 2-dim variable set 
in general.

(i)  Take $\bm{x}_{T}=\bm{T}$ and 
$\bm{y}_{T}=\bm{\not \!\!E}_T$, if no jet is involved in the process, 

(ii) If jets are involved in the process, the fluctuation of 
$\bm{\not \!\!E}_T$ is strongly correlated with that of the jet energy. 
In this case,   take
\begin{equation}
\bm{x}_{T}=\bm{T}+\sum_{j=1}^{N_{jet}}\bm{c}_{Tj},\:\:
\bm{y}_{T}=\not\!\!\bm{E}_T+\sum_{j=1}^{N_{jet}}\bm{E}_{Tj}
\label{eq:metxy}
\end{equation} 
where  $j$ is the jet number,  $\bm{E}_{Tj}$ and $\bm{c}_{Tj}$ are 
the jet and corresponding parton transverse energy, respectively. 
Parton transverse momenta $\bm{c}_{Tj}$'s are independently inferred from 
jets. 

In both cases, we assume  ($\bm{x}_{T},\bm{y}_{T}$) part of the transfer 
function can be factored out as
\begin{equation}
\label{eq:dumis}
	w(\bm{y}|\bm{x}) \propto w (\bm{y}_{T}|\bm{x}_{T}),
\end{equation}
with a normalization condition,
\begin{equation}
\label{eq:normmet}
\int^{\bm{y}_{Tmax}}_{\bm{y}_{Tmin}}w(\bm{y}_{T}|\bm{x}_{T})d^2\bm{y}_{T} = 1.
\end{equation}
Inference of $\bm{x}_{T}$ is made by a 2-dim n.u.r.n. as 
\begin{equation}
\label{eq:unimet} 
 \bm{u} = 
\int^{\bm{x}_{T}}_{\bm{x}_{Tmin}} w(\bm{y}_{T}|\bm{x}) d\bm{x} \Bigg/ 
     \int^{\bm{x}_{Tmax}}_{\bm{x}_{Tmin}} w(\bm{y}_{T}|\bm{x}) d\bm{x}.
\end{equation}

\subsection{ Inference and use of $\bm{s}_{r}$}

\underline{\it Inference of $s_{s}$}$\:\:$
We consider a case where resonance $r$ is selected as a primary parton  
and $s_{r}$ is inferred with the normalized propagator factor,  
$\Pi_{N}(s_{r})=N \Pi(s_{r})$, as a p.d.f. for $s_{r}$, i.e.
\begin{eqnarray}
\int_{0}^{\infty}\Pi_{N}(s_{r})ds_{r}
&=& N \int_{0}^{\infty}\Pi(s_{r})ds_{r}=1,\\
N &=& 1\Big/ \int_{0}^{\infty}\Pi(s_{r})ds_{r}
\approx \frac{M\Gamma}{\pi}.
\end{eqnarray}
Multiplying $\delta(s_{r}-( \sum_{i=1}^{j} c_{i} )^{2})\; ds_{r}(\,=1\,)$ 
to Eq. (\ref{eq:dsdfi}), one gets
\begin{eqnarray}
\label{eq:dsdfi1}
\frac{d\sigma}{d\Phi_{n}^{(f)}} &=& 
\left[ \frac{d\sigma}{d\Phi_{n}^{(f)}}\right]_{c}
\delta(s_{r}-( \sum_{i=1}^{j} c_{i} )^{2})ds_{r},\\
&=& \left[ \frac{d\sigma}{d\Phi_{n}^{(f)}} \right]_{c}
\delta (u-\bar{u}) du, 
\label{eq:dsdfi2}
\end{eqnarray}
where
\begin{eqnarray}
\label{eq:usr}
u(s_{r}) & \equiv & \int_{0}^{s_{r}} \Pi_{N}(s)ds= 
\int_{0}^{s_{r}}\Pi (s) ds \Bigg/ 
\int_{0}^{\infty}\Pi (s) ds,\\
\label{eq:du}
du &=& \Pi_{N}(s_{r}) ds_{r},\\
\bar{u} &\equiv& u(( \sum_{i=1}^{j} c_{i} )^{2}).
\end{eqnarray}
Equation (\ref{eq:du}) indicates $\Pi_{N}(s)$ is a p.d.f. for $s_{r}$.
Thus, in the reconstruction,  
scanning of $s_{r}$ can be made efficiently by generating a n.u.r.n. 
$u\;(0<u<1)$, and making a variable transformation from $u$ to $s_{r}$ by
Eq. (\ref{eq:usr}).

If there are a total of $n_{r}$ resonances,  
$|{\cal{M}}(a+b\rightarrow C)|^{2}$ contains $n_{r}$ propagator factors, and 
one can choose $h$ $(h\le n_{r})$ resonances as primary partons. 

The d.c.s. in this case is written as 
\begin{equation}
\label{eq:dsdfimul}
\frac{d\sigma}{d\Phi_{n}^{(f)}} = \frac{d\sigma}{d\Phi_{n}^{(f)}}
\left[
\prod_{r=1}^{h} \delta(s_{r}-( \sum_{i=1}^{j_{\rho}} 
c^{(\rho)}_{i} )^{2})ds_{r}
\right].
\end{equation}
The values of $s_{r}$ is inferred with $\Pi (s_{r})$ ($r=1,\cdots,h$), 
and $h$ components of daughters, one for each $r$, are determined by solving 
simultaneous equations, 
\begin{equation}
\label{eq:multiqsq}
s_{r}-( \sum_{i=1}^{j_{\rho}} c^{(\rho)}_{i})^{2}=0\quad\quad
(r=1,\cdots, h).
\end{equation}
The value of $d\sigma/d\Phi_{n}^{(f)}$ is to be evaluated using  
momentum components thus determined.

For multiple resonances, $s_{r}$'s can be scanned independently by 
Eq. (\ref{eq:du}).

\underline{\it $\bm{s}_{r}$ for observable daughters}$\:\:$
When daughters of a resonance are all observable, one can evaluate $s_{r}$ 
by Eq. (\ref{eq:qsq}), using $\bm{c}_{i}$'s inferred with T.F. and assumed 
masses  of the final partons.

An alternative way of reconstruction is to infer $s_{r}$ according to  
Eq. (\ref{eq:du}). This is more efficient 
than scanning daughter momenta $c_{i}$'s independently,   
because independent scanning of $c_{i}$'s generally results in 
off-resonant value of $s_{r}$. 

An example is process $W\rightarrow q\bar{q'}$. 
We assume that directions of 2 partons are regenerated from those of 2 jets 
with their T.F., and ask energies of 2 jets. In this case,  
one regenerates $s_W$ by Eq. (\ref{eq:du}) and the energy of  
one parton by Eq. (\ref{eq:uniobs}), 
then the energy of the other parton is given  by solving equation  
$s_W = (q+\bar{q'})^2$, and its T.F. is used as a factor of the likelihood.

\underline{\it $\bm{s}_{r}$ for missing partons} $\:\:$
We consider a process, where there are $m$ missing partons,  
$\not \!\!c_{1},\cdots,$ $\not \!\!c_{m}$, 
and $n_{r}$ intermediate partons. The degree of freedom for missing partons is
$3m$, while measurement of $\not \!\! \bm{E}_{T}$ gives two constraints. Thus, 
if $n_{r} \ge 3m-2$, 
one regenerates $\bm{s}(s_1, \cdots,s_h)(h=3m-2)$ using Eq. (\ref{eq:du})  
and solves Eq. (\ref{eq:multiqsq})  for $\not \!\! \bm{c}$. Then all 
components of  $\not \!\! \bm{c}$ are determined. 

If $n_r < 3m-2$,  the degree  of freedom for $\not \!\bm{c}$ is
\begin{equation}
d = 3m-2-n_{r}>0, 
\end{equation}
and $d$ components of missing partons remain undetermined. 

\underline{\it Examples of d=0 case}$\:\:$ Examples of 1 and 2 missing 
particles are given in the following. 

Example 1: Single $W\rightarrow l\nu$ production associated with/without jets.
In this case, $m=1(\nu )$, $n_r =1(W)$, hence if we 
regenerate $s_W$, then $h=1, d=0$, and Eqs. (\ref{eq:qsq}) and 
(\ref{eq:tmis}) lead to a quadratic equation for $\nu_z$. 
The parton kinematics is determined within two-fold ambiguity. 

Example 2: Dilepton channel in $t\bar{t}$ production, 
\begin{equation}
t\bar{t} \rightarrow l^+ l^- b \bar{b} \nu \bar{\nu}.
\end{equation}
For this process, $m=2$ ($\nu$ and $\bar{\nu}$), and 
$n_r = 4 (t, \bar{t}, W^+, W^- )$, hence
$d=0$, if we regenerate $s_t, s_{\bar{t}}, s_{W^+}, s_{W^-}$ 
by propagator factors and $\bm{T}$ by the transfer function.
Six constraints by Eqs. (\ref{eq:qsq}) and (\ref{eq:tmis}) 
lead to a bi-quadratic equation for $E_{\nu}$ and $E_{\bar{\nu}}$, and 
the parton kinematics is determined within 4-fold ambiguity\cite{kk3}.  


\underline{\it Undetermined variables of  missing partons($d>0$)}
$\:\:$There are cases where some components of the parton 
momenta are left undetermined ($d>0$): 
e.g.   in search for SUSY particles where many missing particles are
involved in the process. If a parton momentum contains such component(s), the 
parton is to be assigned as primary, and the component(s) are to be scanned
uniformly in the phase space.\\ 

Example 3: Charged Higgs production in $t\bar{t}$ channel
\begin{eqnarray}
t\bar{t} \rightarrow (bW^+ ) ( \bar{b} H^-) 
&\rightarrow&
(b l^+ \nu) (\bar{b} \tau^- {\bar{\nu}}_{\tau})\\
&\rightarrow& 
(b l^+ \nu) (\bar{b} l^- {\bar{\nu}}_l \nu_l {\bar{\nu}}_{\tau}),
\end{eqnarray}
where $l=e$ or $\mu$. Here, $m=3({\bar{\nu}}_l, \nu_l, {\bar{\nu}}_{\tau})$, 
and $n_{r}  = 5$  $(t, \bar{t}, W^+ , H^- , \tau^-)$, hence with $h=5$, $d=2$. 
To determine the kinematics, one regenerates  
$s_t, s_{\bar{t}}, s_{W^+}, s_{H^-}, s_{\tau}$ with the propagator factors, 
$T_x, T_y$ with the transfer functions, and  
any 2 ($=d$) components of neutrino momenta uniformly in the phase space.  
Then neutrino equations are reduced to the case of Example 2. 

%

\section{Likelihood of Reconstructed Paths}

\subsection{Likelihood for a single path and multiple paths  
in an event} 
In this subsection, formulas are for each set of 
$(i_{p}, i_{t}, i_{s})$, which are abbreviated. 

\subsubsection{Luminosity factor and the likelihood for real events}
To infer a set of single path kinematics, $\bm{P}(P_{1},\cdots,P_{n})$, for a 
given event, we use in general  virtual mass squared of resonances 
$\bm{s}_{r}$ and parton kinematic variables $\bm{x}$. For a set of $\bm{P}$,  
we define 
the likelihood of the path similar to $L_{1}^{(0)}$ of Eq. (\ref{eq:likemu}).
The only modification required for real data is that for event detection 
efficiency (acceptance). Denoting the efficiency by $\epsilon(\varsigma)$, 
the luminosity factor $l_{1}$ is to be replaced with 
\begin{equation}
\tilde{l}_{1}=\frac{2\; L_{int}}{N_{tot}}
=\frac{2}{\epsilon(\varsigma) \sigma_{T}}.
\end{equation}
and by replacing $l_{1}$ with $\tilde{l}_{1}$, the expected number of events 
has the same form, 
\begin{equation}
\tilde{\mu}=\tilde{l}_{1}\;\frac{d\sigma}{d\Phi_{n}^{(f)}}.
\end{equation}
The likelihood for path $k$ in event $i$ is given by  
\begin{equation}
L_{1}(\bm{\alpha}, \bm{P} | \bm{y}||i,k)
= L_{1}^{(0)}(\tilde{\mu}_{ik})
\label{eq:splkhd},
\end{equation}
if all components of $\bm{x}$ are used to define $\bm{P}$. 
If $\bm{P}$ is defined with 
unused components of $\bm{x}$, $\bm{x}'$, the T.F. for these components 
is to be multiplied to likelihood $L_{1}(\mu)$, namely
\begin{equation}
L_{1}(\bm{\alpha}, \bm{P} | \bm{y}||i,k)
=  L_{1}^{(0)}(\mu_{ik})\: w(\bm{y}'|\bm{x}'||\bm{\alpha},i,k)
\end{equation}

\subsubsection{Likelihood for multiple paths in an event}
To infer the unknown true path of an event, one makes multiple 
path reconstructions. 
Here we discuss three kinds of the likelihood for the true path. 
The advantage of one to the others depends on the process and 
the purpose of analysis. 

\underline{\it Maximum likelihood}$\:\:$
The M.L.E. of $\bm{x}$, $\bm{P}$ and $\bm{\alpha}$ in an event, 
which we denote by $\hat{\bm{x}}$, $\hat{\bm{P}}$ and 
$\hat{\bm{\alpha}}_{1}$, are 
obtained by (a) using general purpose minimum search programs for 
$-2ln(L_{1})$ or by (2) joint likelihood for multiple paths in an event.
to be discussed in the following. 
By the use of  $\hat{\bm{P}}$ and $\hat{\bm{x}}$,  
one can define a likelihood for the $i$-th event, 
as a function of $\bm{\alpha}$, 
$L_{1}(\bm{\alpha}| \bm{y}, \hat{\bm{P}}||i)$.

\underline{\it Expectation value of likelihood}$\:\:$
The expectation value of the likelihood for $\bm{\alpha}$ as 
obtained by a total of $K$ paths for the $i$-th event is defined 
by
\begin{equation}
\overline{L}_{1} (\bm{\alpha}|\bm{y}||i) 
= \frac{1}{K} \sum_{k=1}^{K} 
              L_{1}(\bm{\alpha},\bm{P}| \bm{y}||i,k)\\ 
\label{eq:lpathexp}
\end{equation}


The expectation values of $\bm{x}, \bm{s}_{r}, \bm{P}$ and $\bm{\alpha}$,  
which we denote by $\overline{\bm{x}}$, $\overline{\bm{s}}_{r}$, 
$\overline{\bm{P}}$ and $\overline{\bm{\alpha}}_{1}$, are obtained 
as their means weighted by $L_{1}(\bm{P},\bm{\alpha} | \bm{y}||i,k)$. 

\underline{\it Joint likelihood}$\:\:$
The value of true value of parton kinematics $\bm{P}$, $\bm{P}_{0}$, 
in an event is unknown but common to all 
reconstructed paths in an event, namely,  $\bm{P}_{0}$ is identified with a 
parameter set. 
Reconstructions of $\bm{P}$ can thus be interpreted as pseudo-experiments 
to determine $\bm{P}_{0}$, 
where the single path likelihood plays a role of p.d.f. for $\bm{P}$.
Formally, one inserts $\delta(\bm{P}-\bm{P}_{0})dP$ $(=1)$ into the 
likelihood, interpreting $\bm{P}$ and $\bm{P}_{0}$ as variable and 
parameter sets, respectively.  
The joint likelihood for $K$ paths, 
\begin{equation}
L_{1}^{(K)}(\bm{\alpha},\bm{P}|\bm{y}||i) 
\equiv \left[ \prod_{k=1}^{K} 
             L_{1}(\bm{\alpha}, \bm{P} | \bm{y}||i,k) \right]^{1/K},  
\label{eq:1evjoint}
\end{equation}
can be used to get  the M.L.E. $\hat{\bm{P}}$ 
by the method of maximum likelihood (m.m.l.)\cite{cramer}. 
The likelihood as a function of $\bm{\alpha}$ with $\hat{\bm{P}}$ obtained 
from the joint likelihood is denoted by 
$L_{1}^{(K)}(\bm{\alpha}| \bm{y}, \hat{\bm{P}}||i)$.

\subsection{Likelihood for process, topology and solution}
In the preceding subsection, the likelihood is for a given set of 
$(i_{p}, i_{t}, i_{s})$ in an event. We consider next the use of DLM for 
selection of these integers. 

\underline{\it Integer likelihood $\Lambda(i_{p},i_{t},i_{s})$}$\:\:$
We denote the likelihood for these integers by $\Lambda(i_{p},i_{t},i_{s})$. 
The integer likelihood can be normalized as
\begin{equation}
\sum_{i=1}^{N_{p}} \sum_{j=1}^{N_{t}} \sum_{k=1}^{N_{s}} 
\Lambda(i,j,k) = 1.
\end{equation}

Individual likelihoods for $i_{p}$,  $i_{t}$ and  $i_{s}$ are given by   
\begin{eqnarray}
\label{eq:lambdap}
\Lambda_{p} 
&\equiv& \sum_{i_{t}} \sum_{i_{s}} \Lambda(i_{p},i_{t},i_{s}),\\ 
\label{eq:lambdat}
\Lambda_{t} (i_{p}) 
&\equiv&\sum_{i_{s}}  \Lambda(i_{p},i_{t},i_{s}),\\ 
\Lambda_{s}(i_{p}, i_{t}) &\equiv& \Lambda(i_{p},i_{t},i_{s}). 
\label{eq:lambdas}
\end{eqnarray}
Likelihood $\Lambda_{p}$ is used to discriminate the background against the 
signal, $\Lambda_{t} (i_{p})$ to select topology in a signal-like event, and 
$\Lambda_{s}(i_{p}, i_{t})$ to choose solution of Eq. (\ref{eq:qsq}) for a 
likely topology in the signal-like event.

\underline{\it Evaluation of $\Lambda(i_{p},i_{t},i_{s})$ by DLM}
$\:\:$
The values of $\Lambda$'s are often provided from 
other information, e.g. b-tagging with vertex measurement selects  
certain processes and topologies.
We denote $\Lambda$'s from the other information by $\Lambda^{(0)}$'s, and 
define the integer likelihood as a function of $\bm{\alpha}$ 
for the $i$-th event by
\begin{equation}
\label{eq:l1star}
L^{*}_{1}(i_{p}, i_{t}, i_{s},\bm{\alpha}|\bm{y}||i)
=\Lambda^{(0)} (i_{p}, i_{t}, i_{s})
\lambda_{i}(\bm{\alpha})_{i_{p}, i_{t}, i_{s}} ,
\end{equation}
where $\lambda_{i}$ is the likelihood for the multiple 
inferences in an event as  defined in the preceding subsection,  
\begin{eqnarray}
\nonumber
\lambda_{i}(\bm{\alpha})_{i_{p}, i_{t}, i_{s}}
\:&=&
\:L_{1}^{(x)}(\bm{\alpha}|\bm{y},\hat{\bm{P}}\,||i_{p}, i_{t}, i_{s},i),\quad
\overline{L}_{1} (\bm{\alpha}|\bm{y}||i_{p}, i_{t}, i_{s},i)\quad  \\
&or& 
\quad L_{1}^{(K)}(\bm{\alpha}|\bm{y},\hat{\bm{P}}\,||i_{p}, i_{t}, i_{s},i).
\label{eq:smlambda}
\end{eqnarray}  

The values of likelihood $L_{1}^{*}$'s defined by Eq. (\ref{eq:l1star}) and 
(\ref{eq:smlambda}) are functions of $\bm{\alpha}$. Thus it is appropriate 
to take  their mean value in the search range of $\bm{\alpha}$. 
 Denoting their mean  values by $\overline{L}_{1}^{*}$, 
$\Lambda$'s are given by
\begin{equation}
\nonumber
\Lambda(i_{p}, i_{t}, i_{s})
= \frac{ \overline{L}_{1}^{*}} 
{\sum_{i_{p}} \sum_{i_{t}} \sum_{i_{s}}
\overline{L}_{1}^{*}}
\Bigg|_{[\bm{\alpha}]}.
\label{eq:lambdasbyl}
\end{equation} 
where  suffix $[\bm{\alpha}]$ stands for the search 
region. 

If the search region is wide, 
the discrimination power for  
$(i_{p}, i_{t}, i_{s})$ is weak. Thus  
evaluation of $\Lambda$'s and squeezing the search region of $\bm{\alpha}$  
are to be alternately  iterated. The M.L.E. of $\bm{\alpha}$ is obtained 
by using all events in data, as we discuss in the next subsection. 
If values of $\Lambda$'s converge after the iterations, statistical 
selection of $i_{p}$, $i_{t}$ and $i_{s}$ can be made. 
The whole procedure studied with Monte 
Carlo events can be applied to real data. 

\subsection{Maximum likelihood estimate of $\bm{\alpha}$ from multiple events}
The determination of $\bm{\alpha}$ is to be made by 
$\hat{\bm{\alpha}}_{Nev}$, i.e. M.L.E. from a total of $N_{ev}$ events 
in the given sample. 
The simplest way is to fit  the distribution of $\hat{\bm{\alpha}}_{1}$ 
for individual events, obtained from Eq. (\ref{eq:mlel1}), with those of 
Monte Carlo events with known values of $\bm{\alpha}$. 
The minimum $\chi^{2}$ of the fit gives $\hat{\bm{\alpha}}_{Nev}$ 
\cite{cdftop}. 

Since events are mutually independent, the  $\hat{\bm{\alpha}}_{Nev}$ 
search can also be made with the joint likelihood of $N_{ev}$ events. 
Namely, $\hat{\bm{\alpha}}_{Nev}$ 
is $\bm{\alpha}$ that maximizes the joint likelihood,
\begin{equation}
\label{eq:lnev}
L_{Nev}(\bm{\alpha})=\prod_{i=1}^{Nev}\: \sum_{(i_{p}, i_{t}, i_{s})} 
L^{*}_{1}(i_{p}, i_{t}, i_{s},\bm{\alpha}|\bm{y}||i),
\end{equation}
with $L_{1}^{*}$ given by Eqs. (\ref{eq:l1star}) and (\ref{eq:smlambda}). 
If the selection of $(i_{p}, i_{t}, i_{s})$ is not uniquely made, 
$\hat{\bm{\alpha}}_{Nev}$ determined from 
Eq. (\ref{eq:lnev}) is generally 
shifted from true value $\bm{\alpha}_{0}$ because of remaining false sets of 
$(i_{p}, i_{t}, i_{s})$ in the sum. 
This deviation is to be corrected by the Monte Carlo simulation. 

As we discussed in the preceding subsection, alternate iterations of 
the $\hat{\bm{\alpha}}_{Nev}$ search and the selection of 
$(i_{p}, i_{t}, i_{s})$ are to be made. If the value of 
$\hat{\bm{\alpha}}_{Nev}$ converges, it  can be used 
to redetermine $\hat{\bm{P}}$ and $L_{1}^{*}$  in each event. 

\section{Summary and Comments}

The dynamical likelihood method ( DLM ) is formulated as a procedure to 
reconstruct the quantum process. 

\underline{\it General comments on the formulation}$\:\:$
For a single event reconstruction, we require 
3 quantum conditions: (1) the d.c.s. is per unit phase space,
$d\sigma/d\Phi_{n}^{(f)}$,  (2) the transfer 
variable spaces are quantized by Jacobian scaled variables, (3) the likelihood 
is defined by the Poisson probability for 1 event. 
In condition (1) the final state density which plays an important role 
in the traditional use of the d.c.s. is missing. 
By the Jacobian scaled variables in condition (2),  
the Jacobian factor, i.e. the final state density, is absorved in the 
quantized path, and  the use of T.F. with ordinary quantites are justified. 
The state density is resumed implicitely 
by condition (3), since the number of event distribution, which is the outcome 
of the traditional form of d.c.s., is given by the likelihood of our 
definition. In short, one can forget the Jacobian factor in the formulation 
given in this paper. Only exception is the totally missing particles, the 
reconstruction of which should be made per unit phase space. The integration 
by unknown variables is not to be made in this formulation.

The luminosity factor $l_{1}$ is a constant depending on the event detection 
efficiency. This factor can be obtained from the mean value of the d.c.s 
( for the reconstructed parton kinematics ) of individual events,  
the integrated luminosity and the total number of candidate events. 
In this formulation, the absolute value of the likelihood, i.e. the coupling 
constant for the process, and the dynamical parameters are simultaneously 
determined.

The formulation is more suitable than the earlier ones, Refs. \cite{kk1}$\sim$
\cite{d0}, to analyse events of the collider experiments with 4$\pi$ detectors.

\underline{\it Procedure of path reconstruction}$\:\:$ 
Given a set of observables of an event, one defines the primary partons and  
infers a path. A path is sorted by the physics process, the 
parton-observable identification (topology) and the solution for the momentum 
components of the secondary daughter partons.  

Dynamical constants and parton kinematics in a path are inferred by random 
number generations: (a) dynamical parameters uniformly,    
(b) 3-momenta of observable primary partons according to transfer functions,  
and/or 
(c) virtual masses of intermediate partons with propagator 
factors. If there remain undetermined momentum components of missing partons, 
(d) they are to be inferred uniformly in the phase volume of the partons.  

\underline{\it Applications}$\:\:$
Selections of the process, the topology, and the solution for momentum 
components of the secondary partons, which are specified by integers,  are 
made by the likelihood values for multiple inferences in an event. 
The parton kinematics for each event is given by the M.L.E. or the 
expectation value in the event. 
Dynamical parameters are given by the M.L.E. from the joint likelihood 
of all events. 
Iterations with alternate evaluation of  the likelihood for the integers and 
for the continuous  variables/parameters are  important.\\  
 
Finally, we comment on the use of DLM for new particle searches. 
Most theoretical models of new particles provide  
forms of the d.c.s. that can be used for DLM. 
In addition, the mass value does not strongly depend  on details of the 
parton dynamics, but only on its essential part, i.e. the propagator 
factor of the particle in search. 
Thus the search for theoretically unpredicted new particles by DLM  
is also made possible. 

\section{Acknowledgment}
The author would like to express his thanks to Drs. Alvin V. Tollestrup, 
Melvyn E. Shochet and other members of the CDF (Collider Detector at Fermilab) 
collaboration. This paper came out from useful discussions with the 
collaborators. 

Thanks are also to Drs. Y. Hara, I. Ohba, S. Parke and M. Mangano for 
suggestions and comments during the course of development of the DLM 
formulation.
K. Yorita, K. Ebina, R. Tsuchiya, Y. Kusakabe, J.  Naganoma, S. Toya, 
K. Ikado and T. Arisawa have made examinations of the method in specific 
processes, which are crucial  for the present formulation.

\end{document}